# НЕЯВНАЯ АДАПТАЦИЯ СЕТОЧНОЙ МОДЕЛИ НЕСТАЦИОНАРНОЙ ТЕПЛОПРОВОДНОСТИ К НАГРЕВАЕМОМУ ВЕЩЕСТВУ

## Жуков П.И.[1], Фомин А.В.[2]

*(Старооскольский технологический институт им. А.А. Угарова (филиал) НИТУ «МИСиС», Старый Оскол)*

## Глущенко А.И.[3]

*(ФГБУН Институт проблем управления им. В.А. Трапезникова РАН, Москва)*

*Аннотация. Уравнения нестационарной теплопроводности при моделировании нагрева в высоких температурах требуют проведения процедуры адаптации динамики изменения теплофизических коэффициентов от температуры под конкретную физико-химическую конфигурацию нагреваемого вещества. Решение поставленной проблемы чаще всего осуществляется путем аппроксимации табличных замеров искомых параметров, известных в литературе, при помощи регрессионных уравнений. Такой подход не лишен неустранимой погрешности, которая находится в самих замерах, и адекватно оценить которую затруднительно. Кроме того, для ситуаций тепловой обработки стали применение такого подхода осложняется отсутствием табличных дискретных замеров для большинства неоднородных конфигураций вещества (например, для легированных сталей). В рамках работы предлагается подход, основанный на решении смежной вариационной задачи, основная идея которого заключается в замене процесса адаптации в классическом понимании (нахождение тепловых зависимостей теплофизических параметров от температуры) на «обучение с учителем» сеточной модели по технологическим данным с реального агрегата. Используя градиентный метод, получены формулы для настройки параметров модели нестационарной теплопроводности, отвечающих за теплофизические коэффициенты. Предполагая, что решение смежной вариационной задачи позволит получить неявным образом адаптированную модель, был проведен численный эксперимент для сталей конкретной группы марок, по которой имеются в достаточном количестве как технологические данные, так и табличные. В результате «обученная» сеточная модель, не получавшая явным образом никаких сведений о физико-химической конфигурации вещества, показала среднюю ошибку 18,82 °C, что незначительно больше средней ошибки модели, адаптированной классическим образом по табличным данным (18.1 °C).*

---

[1] *Жуков Пётр Игоревич, аспирант (Zhukov.petr86@yandex.ru)*

[2] *Фомин Андрей Вячеславович, к.т.н., доцент (verner444@yandex.ru)*

[3] *Глущенко Антон Игоревич, д.т.н., в.н.с. (aiglush@ipu.ru)*

## 1. Введение

На сегодняшний день цифровизация существующих промышленных производств приобретает все более широкий характер [15, 24]. Внедрение ее идей приводит к росту популярности неклассических подходов к управлению технологическими агрегатами, таких как, например, Advanced Process Control (APC) [22] или Model Predictive Control (MPC) [18], а также к развитию такой концепции как «цифровой двойник» или «цифровая тень» [19]. Все эти методы и подходы базируются на обработке большого количества данных, собранных с объекта автоматизированными системами управления (АСУ), с целью получить как можно более точные модели протекающих в нем базовых процессов и подпроцессов.

Основное функциональное преимущество моделей, построенных в рамках указанных выше подходов, перед классическими математическими моделями заключается в их прогностической способности. Такие цифровые модели позволяют рассматривать жизненный цикл технологического объекта или процесса с опережением реального времени, что позволяет быстро и оперативно принимать корректирующие решения.

Одним из перспективных направлений приложения вышеизложенных идей является температурная обработка стали в промышленных тепловых установках (печах). В машиностроении и металлургии высокотемпературный нагрев стальных заготовок применяется очень широко, при этом данный процесс является крайне энергоемким и сложным для контроля [21]. Для металлургического производства данная проблема особенно актуальна на сегодняшний день, так как потери металла при нагреве могут составлять до 40% себестоимости нагрева, а затраты на топливо до 20% [16]. Сложность построения энергоэффективных систем управления для высокотемпературных тепловых агрегатов связана, в первую очередь, с комплексностью внутрипечных факторов, невозможности точно оценить их комбинаторные влияния друг на друга.

На сегодняшний день, управление такими агрегатами ведется по факторам, которые могут быть измерены или идентифицированы с достаточной степенью точности. Например, контроль за нагревом стальных заготовок в печи ведется по температуре внутрипечной атмосферы, а не по температуре поверхности заготовки. Сложности добавляет также высокая инерционность тепловых процессов, из-за которой принятие решений, например, об изменении задания на локальные контура управления температурой в зонах печи, необходимо осуществлять заранее. Кроме того, это решение всегда будет ограничено технологической инструкцией по ведению процесса и, в случае некорректного или несвоевременного воздействия, моментально устранить последствия будет невозможно. Таким образом, необходимо предвидеть результат применения таких решений.

Вышеизложенное создает потребность в моделях, способных точно описать тепловые процессы и, опережая реальное время, прогнозировать «выход» процесса нагрева (финальную температуру заготовок) еще до того, как он сам будет завершен. На такую модель, в том числе, может опираться предлагаемая в работе [16] система экстремально-оптимизирующего управления, призванная снизить издержки нагрева.

Проблема моделирования высокотемпературной обработки стали не нова, однако, из-за множества различных нестационарных режимов нагрева (отжиг, отпуск, закалка, методический нагрев и т.д.) единого решения данной задачи нет, а поиск частных решений, как, например, в работах [4,5,14] до сих пор остается актуальным. Современным способом описать нестационарный высокотемпературный нагрев стали является численное решение краевых задач на основе дифференциального уравнения нестационарной теплопроводности [4].

Особенностью данного подхода является его масштабируемость, что хорошо видно на примере работы [3], где модель нестационарной теплопроводности расширяется стационарными тепловыми моделями. При этом, в ситуациях высокотемпературного нагрева точность моделей сильно зависит от выбора значений критически важных параметров, таких как теплоемкость $c$, теплопроводность $\lambda$, плотность стали $\rho$ др.

Сложность проблемы также повышает тот факт, что всю выпускаемую предприятием сталь нельзя рассматривать как единый материал из-за отличий в его физико-химической конфигурации (разных «марок» стали, динамика прогрева которых различается) [17]. При этом, из-за того, что при высоких температурах нельзя пренебречь изменением упомянутых $c$, $\lambda$, $\rho$ материала от температуры, возникает актуальная подзадача «адаптации» условно универсальных математических моделей к нагреваемому веществу. Причем такие изменения индивидуальны для каждой марки стали.

Ранее уже предпринимались попытки адаптировать модель нестационарной теплопроводности к стальным заготовкам разной физико-химической конфигурации [9, 11]. Данная адаптация была основана на аппроксимации дискретных замеров искомых параметров (плотности, теплоемкости и теплопроводности), представленных в литературе, регрессионными функциями. В результате экспериментов было установлено, что для сохранения робастности и точности модели для каждого материала придется проводить отдельную адаптацию. Для описания нагрева стали это становится отдельной проблемой, так как нет истинных формул, по которым можно было бы восстановить температурные зависимости для искомых параметров $c$, $\lambda$, $\rho$ материала. Одновременно с этим, общедоступные таблицы эмпирических дискретных замеров, как например [12], по которым можно было бы восстановить эти зависимости, как правило, являются неполными. Кроме того, подобные «марочники», из числа тех, что есть в открытом доступе часто игнорируют современные «марки» стали. В этих условиях, фактически, точность найденной эмпирически зависимости сильно зависит не столько от неустранимой ошибки метода аппроксимации, сколько от ошибки в наблюдениях.

В рамках данной работы рассматривается задача построения условно универсальной математической модели с особым способом «адаптации» к нагреваемому веществу путем решения вариационной задачи на основе модели нестационарной теплопроводности.

Основа подхода, излагаемого далее, заключается в замене «адаптируемых» непрерывных параметров модели ($\rho$, $\lambda$, $c$) на свободные безразмерные дискретные параметры ($\omega$ и $\varphi$) с поиском, в дальнейшем, их условно оптимальных значений при помощи решения вариационной задачи на основе имеющихся реальных технологических данных с конкретной печи. Предполагается, что подобный подход, основанный на градиентом методе, позволит преодолеть указанные недостатки, в частности нехватку «табличных» замеров для восстановления $c, \lambda, \rho$, и выполнить «привязку» модели к особенностям работы конкретного объекта.

## 2. Исходная модель нагрева в виде конечных разностей

Рассмотрим модель теплового нагрева объектов в нестационарных условиях. В основе такой модели будет лежать дифференциальное уравнение нестационарной теплопроводности, которое заранее определим в прямоугольных координатах $x$ и $y$:

$$(1) \quad \rho \cdot c \frac{\partial T}{\partial t} = \lambda \cdot \left( \frac{\partial^2 T}{\partial x^2} + \frac{\partial^2 T}{\partial y^2} \right),$$

где $\rho$ – плотность, $кг/м^3$; $\lambda$ – теплопроводность, $Вт/(м*К)$; $c$ – теплоемкость, $Дж/(кг*К)$, $T$ – температура нагреваемого материала, $К$. Пусть имеется краевая задача конвективно-радиационного теплообмена с граничными и начальными условиями следующего вида:

$$(2) \quad \begin{cases} t = 0 : T = T_0, 0 \le x \le L, 0 \le y \le H \\ x = 0 : -\lambda \frac{\partial T}{\partial x} = \kappa_1(T_1 - T_{Об}) + \varepsilon\sigma(T_1^4 - T_{Об}^4); \\ x = L : \lambda \frac{\partial T}{\partial x} = q_1, t > 0; \\ y = 0 : -\lambda \frac{\partial T}{\partial y} = q_2, t > 0; \\ y = H : \lambda \frac{\partial T}{\partial y} = \kappa_2(T_2 - T_{Об}) + \varepsilon\sigma(T_2^4 - T_{Об}^4); \end{cases}$$

Здесь на границах тепловое поле формируется на основе уравнения Ньютона-Рихмана и нелинейного условия Стефана – Больцмана:

$$(3) \quad \frac{\partial T}{\partial l} = \kappa \cdot (T_{Cp} - T_{O\delta}) + \varepsilon_{O\delta} \cdot \sigma \cdot (T_{Cp}^4 - T_{O\delta}^4),$$

где $T_{O\delta}$ – температура объекта, $K$; $T_{Cp}$ – температура среды, $K$; $n$ – мерность по пространству; $\kappa$ – коэффициент теплообмена, $Bm/(м^2{*}K)$, $\varepsilon$ – приведенная степень черноты, $\sigma$ – постоянная Стефана-Больцмана, принятая равной $5{,}67{*}10^{-8}$ $Bm{*}м^{-2}{*}K^{-4}$, $T_1$ – температура теплового поля, направленного по нормали вдоль оси $Ox$, $K$; $T_2$ – температура теплового поля, направленного по нормали вдоль оси $Oy$, $K$; $\kappa_1$ и $\kappa_2$ – коэффициенты теплообмена полей с $T_1$ и $T_2$ соответственно, $Bm/(м^2{*}K)$; $\varepsilon_{O\delta}$ – приведенная степень черноты поверхности нагреваемого вещества; $H$ – длина двумерного сечения, $м$; $L$ – ширина двумерного сечения, $м$, $T_0$ – начальная температура нагреваемого объекта, $K$; $q_1$, $q_2$ – тепловые потоки на границах вдоль осей $Ox$ и $Oy$ соответственно, которые в рамках данной задачи принимаются равными нулю, $l$ – это нормаль к поверхности $Ox$ или $Oy$ одномерного поля, которое зависит от выбранного уравнения при рассмотрении системы (2).

Решить краевую задачу возможно методом численного дифференцирования, а конкретно – методом конечных разностей. Для этого предположим, что исходное уравнение (1) определено на двумерном непрерывном пространстве $\bar{G}_{xy} = G_{xy} + \Gamma_{xy}$, где $\Gamma$ – это граница пространства, а $G$ – это внутренняя его часть. Метод конечных разностей предполагает, что $\bar{G}_{xy}$ может быть «замощено» дискретной сеткой $\overline{W}_{xy}$, причем такой, что $\overline{W}_{xy} = w_{xy} + \gamma_{xy}$, где $w_{xy}$ – это дискретная сетка внутреннего пространства $G_{xy}$, а $\gamma_{xy}$ – это дискретная сетка на границе пространства $\Gamma_{xy}$.

Рассмотрим двумерную внутреннюю сетку в следующем виде: $w_{xy} = w_x + w_y$, и, аналогично, граничную сетку: $\gamma_{xy} = \gamma_x + \gamma_y$. Полагаем, что одномерные сетки соответствующих осей равномерны с одинаковым шагом $h$, и имеет место следующая функциональная система:

$$(4) \quad w_{xy} = \begin{cases} w_{hx} = \{ x_k = h_x \cdot k_x \mid k_x = \overline{1, H-1}, h_x \cdot H = x_{\max} \} \\ w_{hy} = \{ y_q = h_y \cdot q_y \mid q_y = \overline{1, W-1}, h_y \cdot W = y_{\max} \} \end{cases}$$

где $h_x$ – это шаг по сетке вдоль $Ox$, а $h_y$ – шаг по сетке вдоль $Oy$, $H$ и $M$ – количество шагов сетки вдоль соответствующих осей,

$x_{max}$ и $y_{max}$ – это границы пространства $\bar{G}_{xy}$ по $Ox$ и $Oy$ соответственно. Аналогично, сетка, спроецированная на $\Gamma_{xy}$, будет иметь вид системы (5):

$$(5) \quad \gamma_{xy} = \begin{cases} \gamma_{hx} = \{ x_k = h_x \cdot k_x \mid k_x = \overline{0, H}, h_x \cdot H = x_{max} \} \\ \gamma_{hy} = \{ y_q = h_y \cdot q_y \mid q_y = \overline{0, W}, h_y \cdot W = y_{max} \} \end{cases}$$

Поскольку уравнение (1) нестационарно по времени, что следует из дифференциального оператора в левой части, то при переходе к конечным разностям потребуется также спроецировать дискретную сетку на ось времени. В данном случае сетка будет одномерной и равномерной и будет иметь вид:

$$(6) \quad w_\tau = \{ n = \tau \cdot k_t \mid k_t = \overline{0, N}, \tau \cdot N = t_{max} \},$$

где $\tau$ – шаг по времени, $N$ – количество узлов сетки по времени, $t_{max}$ – максимальный временной горизонт.

В таком рассмотрении имеется возможность явно спроецировать граничные условия на граничную сетку (5) и при помощи них неявно вычислить значение в узлах внутренней сетки (4) для правой части уравнения (1) и узлах временной сетки (6) для левой части.

Произведя замену методом конечных разностей и применив покоординатное расщепление [7], получим две конечно-разностные системы:

$$(7) \quad \begin{cases} \rho \cdot c \dfrac{T_{x_k, y_q = const}^{n+\frac{1}{2}} - T_{x_k, y_q = const}^{n}}{\tau} = \lambda \cdot \left( \dfrac{T_{x_k+1}^{n+\frac{1}{2}} - 2 \cdot T_{x_k}^{n+\frac{1}{2}} + T_{x_k-1}^{n+\frac{1}{2}}}{h_x^2} \right) \\ \rho \cdot c \dfrac{T_{x_k = const, y_q}^{n+1} - T_{x_k = const, y_q}^{n+\frac{1}{2}}}{\tau} = \lambda \cdot \left( \dfrac{T_{y_q+1}^{n+1} - 2 \cdot T_{y_q}^{n+1} + T_{y_q-1}^{n+1}}{h_y^2} \right) \end{cases},$$

где $n$ – текущий временной слой, $T^{n+1}$ – температура на целом временном слое, $T^{n+1/2}$ – температура на половинчатом временном слое. Температуры по целому $n+1$-му временному слою рассчитываются для узлов с зафиксированной координатой $x$ вдоль оси $Oy$ ($T_{yq}$). Температуры на половинчатом временном слое $n+1/2$ рассчитываются с зафиксированной координатой $y$ вдоль оси $Ox$ ($T_{xk}$). $h_x$, $h_y$ – шаги по пространству для сетки $Ox$ и $Oy$ соответственно, для которых граничные условия, согласно задаче (2), будут иметь также конечно-разностный вид:

$$(8) \quad \begin{cases} -\lambda \dfrac{T^{n+1}_{x_{k+1}} - T^{n+1}_{x_k}}{h_x} = \kappa_1 \cdot (T_1 - T^{n+1}_{x_k}) + \varepsilon \cdot \sigma \cdot ((T_1)^4 - (T^{n+1}_{x_k})^4) \\ \lambda \dfrac{T^{n+1}_{y_{q+1}} - T^{n+1}_{y_q}}{h_y} = \kappa_2 \cdot (T_2 - T^{n+1}_{y_q}) + \varepsilon \cdot \sigma \cdot ((T_2)^4 - (T^{n+1}_{y_q})^4) \end{cases}$$

Решить задачу нестационарной теплопроводности в виде неявных локально-одномерных схем (7) с граничными условиями (8) можно при помощи методов прямой и обратной прогонки, который описан, например в [13] или в более общем виде в [8]. Для этого сначала применим метод трехдиагональной матрицы, согласно которому решение может быть представлено в виде:

$(9) \quad F_l = A_l \cdot T^{n+1}_{l+1} - B_l \cdot T^{n+1}_l + C_l \cdot T^{n+1}_{l-1}$,

где $A$, $B$, $C$ – это элементы трехдиагональной матрицы Якоби, $F$ – это вектор-столбец коэффициентов, связывающих аппроксимацию по трем точкам $T_{l+1}$, $T_l$ и $T_{l-1}$ на последующем временном слое ($n+1$) с $T_l$ на предыдущем ($n$). В дальнейшем индекс $l$ следует трактовать следующим образом: на его место может быть подставлен индекс «$x_k$» или «$y_q$» в зависимости от того о нормали к какой поверхности идет речь ($Ox$ или $Oy$).Например, если рассматривается поверхность $Ox$, то $l$ – это конкретная точка $x_k$ из выбранной локально одномерной пространственной сетки, $l+1$ соответствует $x_{k+1}$, а $l-1 - x_{k-1}$. Также и для поверхности $Oy$. Для рассматриваемой модели данные коэффициенты находятся следующим образом:

$$(10) \quad \begin{cases} A_l = C_l = \dfrac{\lambda}{h_l^2} \\ B_l = \dfrac{2 \cdot \lambda}{h_l^2} + \dfrac{\rho \cdot c}{\tau}, \\ F_l = -\dfrac{\rho \cdot c}{\tau} \cdot T_l^n \end{cases}$$

где $h_l$ – это шаг по пространству вдоль температурного поля с нормалью $l$.

Данные коэффициенты необходимы для расчета прогоночных коэффициентов по системе:

$$(11) \begin{cases} \alpha_l = \dfrac{A_l}{B_l - C_l \cdot \alpha_{l-1}}, \\ \beta_l = \dfrac{C_l \cdot \beta_{l-1} - F_l}{B_l - C_l \cdot \alpha_{l-1}} \end{cases}$$

которые используются в уравнении прогонки (12), связывающем значение температурного поля в узлах сетки $Ox$ или $Oy$ на $n+1$-м временном слое с температурой в $x_{k+1}$ и $y_{q+1}$ узлах этих же сеток на том же $n+1$-м временном слое:

$$(12) \quad T_l^{n+1} = \alpha_l \cdot T_{l+1}^{n+1} + \beta_l.$$

Согласно метода, сначала прямой прогонкой вычисляются коэффициенты $\alpha$ и $\beta$, а затем, методом обратной прогонки вычисляются значения $T$ на $n+1$-м временном слое. Используя (7), (8) и уравнение (12), зафиксировав одну из координат, можно получить начальные значения прогоночных коэффициентов, для оси $Ox$:

$$(13) \begin{cases} \alpha_{x_0} = \dfrac{2 \cdot \lambda \cdot \tau}{\rho \cdot c \cdot h_x^2 + 2 \cdot \tau \cdot (\lambda + \kappa_1 \cdot h_x)} \\ \beta_{x_0} = \dfrac{\rho \cdot c \cdot h_x^2 \cdot T_0^n + 2 \cdot \tau \cdot \kappa_1 \cdot h_x \cdot T_1}{\rho \cdot c \cdot h_x^2 + 2 \cdot \tau \cdot (\lambda + \kappa_1 \cdot h_x)} + \\ + \dfrac{2 \cdot \tau \cdot \varepsilon_1 \cdot \sigma \cdot h_x \cdot ((T_1)^4 - (T_0^{n+1})^4)}{\rho \cdot c \cdot h_x^2 + 2 \cdot \tau \cdot (\lambda + \kappa_1 \cdot h_x)} \end{cases}$$

и для оси $Oy$:

$$(14) \begin{cases} \alpha_{y_0} = \dfrac{2 \cdot a \cdot \tau}{h_y^2 + 2 \cdot a \cdot \tau} \\ \beta_{y_0} = \dfrac{h_y^2 \cdot T_0^n}{h_y^2 + 2 \cdot a \cdot \tau} + \dfrac{2 \cdot a \cdot \tau \cdot h_y \cdot q_1}{\lambda \cdot (h_y^2 + 2 \cdot a \cdot \tau)} \end{cases}$$

Подробнее с процессом вывода (13) и (14) можно ознакомиться, например, в работе [6].

Здесь $a = \lambda/(c*\rho)$ – коэффициент температуропроводности. Для того, чтобы получить уравнения обратной прогонки, необходимо разложить температуру $(n+1)$-го временного слоя в ряд Тейлора до второго порядка:

$$(15) \quad T_{l-1}^{n+1} = T_l^{n+1} - h_l \cdot \frac{\partial T}{\partial l}\bigg|^{n+1} + \frac{h_l^2 \cdot \rho \cdot c}{2 \cdot \lambda} \cdot \frac{\partial^2 T}{\partial l^2}\bigg|^{n+1} + O(h)$$

Разложение (15) применимо к моделям, описывающим температурное поле вдоль осей $Ox$ и $Oy$. Таким образом, подставляя в (15) в качестве первой производной соответствующее уравнение из системы (8), а в качестве второй – соответствующее уравнение из системы (7), можно получить уравнения для расчета температур, предварительно разрешив его относительно $T_l$. Для температуры в узлах сетки при движении вдоль $Ox$ во время обратного прогона имеем:

$$(16) \quad T_{x_k}^{n+\frac{1}{2}} = \frac{2 \cdot a \cdot \tau \cdot \lambda \cdot \beta_{x_{k-1}} - 2 \cdot a \cdot \tau \cdot h_x \cdot q_2 + h_x^2 \cdot \lambda \cdot T_{x_k}^n}{\lambda \cdot h_x^2 + 2 \cdot a \cdot \tau \cdot \lambda \cdot (1 - \alpha_{x_{k-1}})},$$

А при движении вдоль $Oy$ получим:

$$(17) \quad T_{y_q}^{n+1} = \frac{2 \cdot \lambda \cdot \tau \cdot \beta_{y_{q-1}} + 2 \cdot \tau \cdot \kappa_2 \cdot h_y \cdot T_2 +}{2 \cdot \tau \cdot \lambda \cdot (1 - \alpha y_{q-1}) + 2 \cdot \tau \cdot \kappa_2 \cdot h_y + \rho \cdot c \cdot h_y^2} +$$
$$+ \frac{\rho \cdot c \cdot h_y^2 \cdot T_{y_q}^n + 2 \cdot \tau \cdot \varepsilon_2 \cdot \sigma \cdot h_y((T_2)^4 - (T_{y_q}^n)^4)}{2 \cdot \tau \cdot \lambda \cdot (1 - \alpha_{y_{q-1}}) + 2 \cdot \tau \cdot \kappa_2 \cdot h_y + \rho \cdot c \cdot h_y^2},$$

Поскольку имело место покоординатное расщепление двумерной сетки на локально-одномерные, то движение вдоль оси $Ox$ определяет температуру на половинчатом $n+1/2$ слое, а вдоль оси $Oy$ – на целом $n+1$-м слое.

Используя уравнения (12), и вычисляя на каждом $n+1$-м временном слое значение начальных прогоночных коэффициентов по системам (13) и (14), можно методом прямой итерации получить значение коэффициентов $\alpha$ и $\beta$ для каждого узла локально-одномерной сетки. Затем, опираясь также на уравнение (12), но двигаясь обратной итерацией, от значений границ (16) и (17) можно получить значение температур в каждом из узлов сетки.

Такие численные конечно-разностные модели обладают неустранимой ошибкой порядка $O(h^2) + O(\tau)$, однако, на практике,

неустранимая ошибка такой модели дополняется ошибкой адаптации её теплофизических коэффициентов к нагреваемому веществу. Рассмотрим проблему адаптации детальней.

## 3. Проблема адаптации сеточной модели. Постановка задачи

При моделировании численным методом системы (7) в условиях высоких температур нельзя пренебречь зависимостью теплоемкости ($c$), плотности ($\rho$) и теплопроводности ($\lambda$) от температуры нагреваемого вещества. В общем смысле это подразумевает, что для описания данных параметров имеются некоторые функциональные зависимости:

$$(18) \quad \begin{cases} \lambda = f_1(T), \\ \rho = f_2(T), \\ c = f_3(T). \end{cases}$$

Однако существует проблема определения вида этих зависимостей для сложных по химико-физической структуре материалов, например, сталей различной конфигурации (марки) или композитных материалов. Для подобных случаев нет истинных формул для описания данной зависимости и, в основном, прибегают к эмпирическим или критериальным закономерностям, восстановленным по дискретным замерам искомых параметров, приведенным в известной литературе.

При этом, точность моделей системы (7) в достаточной мере определяется точностью $f_1(T)$, $f_2(T)$, $f_3(T)$, что уже рассматривалось авторами в работе [11]. Проблема выбора функциональных зависимостей системы (18) также поднималась в работе [9].

Формально, проблема восстановления теплофизических коэффициентов из температурных потоков или полей относится к классу обратных задач теплопроводности (ОЗТ), которые направленны на анализ свойств нагреваемого материала. Однако, для увеличения точности прямого решения задачи моделирования тепловых потоков и полей подходы ОЗТ применимы мало. Во-первых, они алгоритмически сложны и требуют несколько алгоритмов численного дифференцирования (интегрирования), как например, в работах [1, 2]. Во-вторых, такие методы [10] опираются на допущения об известности двух функций из трех в системе (18), что не всегда выполняется при моделировании реальных процессов нагрева. В-третьих, популярные подходы к ОЗТ, основанные на решении смежных вариационных задач, применяются в граничных условиях I или II-го рода и их сложно экстраполировать на процессы, протекающие в сложных тепловых агрегатах реальных производств.

Одновременно с этим, подходы с решением вариационных задач при рассмотрении ОЗТ [1,2,10], могут быть полезны для решения проблемы адаптации модели. Рассмотрим основную идею применения методов ОЗТ при решении смежных вариационных задач, предполагая, что ни один из искомых теплофизических параметров неизвестен. Для этого предлагается произвести следующую замену теплофизических параметров на безразмерные величины:

$$(19) \quad \begin{cases} \varphi^{(n)} = \lambda \\ \omega^{(n)} = \rho \cdot c \end{cases},$$

где $n$ – это элементарный отрезок одномерной временной сетки, выполненный с шагом $\tau$. При произведенных заменах (19) сокращается количество восстанавливаемых величин системы (18).

Кроме того, ошибка $\varphi$ напрямую может быть оценена по ошибке $E$ самой модели прямого расчета. В таком случае можно пытаться двигаться по ошибке (20) в сторону условно оптимального значения $\varphi$, и при этом оценка ошибки будет явной.

Таким образом исключается из процесса адаптации необходимость поиска непрерывной функции для термодинамических параметров, заменяя их дискретными параметрами $\varphi$ и $\omega$, определенными в каждом узле временной сетки. Другими словами, предлагается заменить правые части системы (18) на вектор-строки настраиваемых параметров фиксированной длины, равной количеству шагов по времени.

Формально это можно записать в следующем виде:

$$(20) \quad \begin{cases} \omega_l = \left\{ \omega_l^{(n)} \mid n = \overline{0, N} \right\} \\ \varphi_l = \left\{ \varphi_l^{(n)} \mid n = \overline{0, N} \right\} \end{cases},$$

где *N* – индекс последнего элементарного отрезка. Таким образом, остается вместо адаптации по нагреваемому веществу решить вариационную задачу и найти параметры $\varphi$ и $\omega$, опираясь на выход модели прямого расчета температурного поля.

В целом, в данном исследовании предлагается заменить поиск непрерывных тепловых зависимостей теплофизических величин на оптимизацию дискретных параметров, отражающих эту зависимость. Другими словами, предлагается выполнить замену (19) таким образом, чтобы получить массивы параметров (20), покрывающие всё время жизни динамической системы с дискретным временем, которой является модель нестационарной теплопроводности в представлении (7) и (8).

Опираясь на вышеизложенное, ставится задача спроектировать механизмы нахождения этих параметров в соответствии с движением к их условно оптимальным значениям при помощи решения обратной вариационной задачи на основе имеющихся данных о работе технологического агрегата. Рассмотрим предлагаемое решение, основой которого является градиентный метод, детальней.

## 4. Решение вариационной задачи

Поскольку в системе (7) имеются две сеточные модели по осям *Ox* и *Oy*, то, соответственно, параметры $\varphi$ и $\omega$ больше не интерпретируются как теплофизические параметры, и мы можем ввести их для каждой модели в отдельности:

$$(21) \begin{cases} \omega_x = \left\{ \omega_x^{(n)} \mid n = \overline{0, N} \right\} \\ \omega_y = \left\{ \omega_y^{(n)} \mid n = \overline{0, N} \right\} \\ \varphi_x = \left\{ \varphi_x^{(n)} \mid n = \overline{0, N} \right\} \\ \varphi_y = \left\{ \varphi_y^{(n)} \mid n = \overline{0, N} \right\} \end{cases},$$

где *N* – максимальное количество элементарных отрезков по временной сетке, отмеренных с шагом τ. Тогда, для сеточных моделей системы (7) будет справедлива следующая замена:

$$(22) \begin{cases} \omega_x^{(n)} \cdot \dfrac{T_{x_k, y_q=const}^{n+\frac{1}{2}} - T_{x_k, y_q=const}^{n}}{\tau} = \varphi_x^{(n)} \cdot \dfrac{T_{x_k+1}^{n+\frac{1}{2}} - 2 \cdot T_{x_k}^{n+\frac{1}{2}} + T_{x_k-1}^{n+\frac{1}{2}}}{h_x^2}, \\[4mm] \omega_y^{(n)} \cdot \dfrac{T_{x_k=const, y_q}^{n+1} - T_{x_k=const, y_q}^{n+\frac{1}{2}}}{\tau} = \varphi_y^{(n)} \cdot \dfrac{T_{y_q+1}^{n+1} - 2 \cdot T_{y_q}^{n+1} + T_{y_q-1}^{n+1}}{h_y^2}, \end{cases}$$

Пусть каждое уравнение системы (22) есть динамическая система с дискретным временем, определенная на одной и той же временной сетке с шагом τ. Пусть состояниями системы вдоль *Ox* будут $\xi^X$, а состояниям системы вдоль $Oy - \xi^Y$.

Тогда имеется два набора состояний: для системы *Ox* (рис.1) и для системы *Oy* (рис.2).

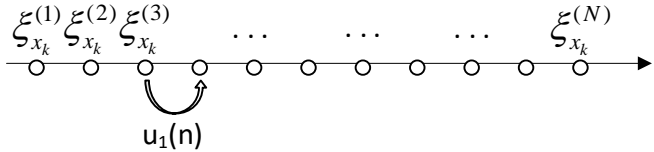

*Рис.1. Набор состояний системы Ox*

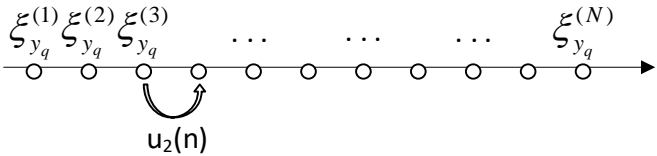

*Рис.2. Набор состояний системы Oy*

Пусть имеются управления $u_1$ и $u_2$. Тогда каждое последующее состояние системы достигается только как рекуррентная функция вида:

$$(23) \begin{cases} \xi_{x_k}^{(n)} = g_1(\xi_{x_k}^{(n-1)}, u_1(n)) \\ \xi_{y_q}^{(n)} = g_2(\xi_{y_q}^{(n-1)}, u_2(n)) \end{cases}$$

Начальное состояние зафиксировано начальными условиями краевой задачи, поэтому для (23) справедливо $n \neq 0$. Таким образом, на основе уравнений системы (22) закономерно будет утверждать, что эти управления $u_1$ и $u_2$ будут представлять собой пару значений настраиваемых параметров $\varphi$ и $\omega$, определенных в каждом временном слое $n$:

$$(24)\quad \begin{cases} u_1(n) = \{\varphi_x^{(n)}, \omega_x^{(n)}\} \\ u_2(n) = \{\varphi_y^{(n)}, \omega_y^{(n)}\} \end{cases}$$

Для того, чтобы получить уравнения состояния каждой модели из системы (22) после сделанных замен (19), произведем соответствующие замены в исходных уравнениях температуры на граничном слое (16) и (17).

Сделав замены (19) в уравнении для определения температуры на половинчатом временном слое (16), получим уравнение состояние динамической системы $Ox$:

$$(25)\quad g_1(T_{x_k}^{(n)}, \{\varphi_x^{(n)}, \omega_x^{(n)}\}) = \frac{2 \cdot (\frac{\varphi_x^{(n)}}{\omega_x^{(n)}}) \cdot \tau \cdot \varphi_x^{(n)} \cdot \beta_{x_{k-1}} - 2 \cdot (\frac{\varphi_x^{(n)}}{\omega_x^{(n)}}) \cdot \tau \cdot h_x \cdot q_2 + h_x^2 \cdot \varphi_x^{(n)} \cdot T_{x_k}^{(n)}}{\varphi_x^{(n)} \cdot h_x^2 + 2 \cdot (\frac{\varphi_x^{(n)}}{\omega_x^{(n)}}) \cdot \tau \cdot \varphi_x^{(n)} \cdot (1 - \alpha_{x_{k-1}})}$$

И для системы $Oy$ соответственно, используя уравнение (17), получим:

$$(26)\quad \begin{aligned} g_2(T_{y_q}^{(n)}, \{\varphi_y^{(n)}, \omega_y^{(n)}\}) &= \frac{2 \cdot \varphi_y^{(n)} \cdot \tau \cdot \beta_{y_{q-1}} + 2 \cdot \tau \cdot \kappa_2 \cdot h_y \cdot T_2}{2 \cdot \tau \cdot \varphi_y^{(n)} \cdot (1 - \alpha_{y_{q-1}}) + 2 \cdot \tau \cdot \kappa_2 \cdot h_y + \omega_y^{(n)} \cdot h_y^2} + \\ &+ \frac{\omega_y^{(n)} \cdot h_y^2 \cdot T_{y_q}^{(n)} + 2 \cdot \tau \cdot \varepsilon_2 \cdot \sigma \cdot h_y((T_2)^4 - (T_{y_q}^{(n)}))^4}{2 \cdot \tau \cdot \varphi_y^{(n)} \cdot (1 - \alpha_{y_{q-1}}) + 2 \cdot \tau \cdot \kappa_2 \cdot h_y + \omega_y^{(n)} \cdot h_y^2} \end{aligned}$$

Опираясь на (25) и (26), можно заметить, что смена состояний динамических систем зависит от значений введенных параметров, которые в виде системы (24) представляют собой потоки управления.

Пусть системы развиваются во времени до $N$, причем в $N$ ожидается определенная эталонная температура $y$, а на вход подается целое количество наблюдений $M$. Имея данные о целевой температуре, а также информацию о времени нагрева ($t_{max}$) и температурах сред ($T_1$ и $T_2$), можно «обучить» сеточную модель на заранее определённой выборке, пытаясь обобщить динамику нагрева различных марок сталей, тем самым избежав надобности в явной адаптации правых частей системы (18).

Рассмотрим процесс «обучения» сеточной модели более детально. Пусть имеется целевые функции для системы $Ox$ (27) и $Oy$ – (28).

(27) $E_x = \dfrac{1}{2} \sum_{i=1}^{M} (y_i - g_1(T_{x_k}^{(n)}, u_1(n)_i)^2$

(28) $E_y = \dfrac{1}{2} \sum_{i=1}^{M} (y_i - g_2(T_{y_q}^{(n)}, u_2(n))_i)^2$

Таким образом, имеет место следующая постановка задачи оптимизации: необходимо найти такие траектории развития системы по $Ox$ и $Oy$, чтобы их целевые критерии минимизировались:

(29) $\begin{cases} \{u_1(n), u_1(n+1), ..., u_1(N)\}, \ \dfrac{1}{2} \sum_{i=1}^{M} (y_i - g_1(T_{x_k}^{(n)}, u_1(n)_i)^2 \to \min \\ \{u_2(n), u_2(n+1), ..., u_2(N)\}, \ \dfrac{1}{2} \sum_{i=1}^{M} (y_i - g_2(T_{y_q}^{(n)}, u_2(n))_i)^2 \to \min \end{cases}$

Пусть все свободные параметры системы инициализируются случайным образом с условием $\omega \neq 0$. Тогда будут получены определенные траектории $u_1(n)$ и $u_2(n)$, которые будут отличны от оптимальных. Явным образом оценить оптимальность управляющих воздействий затруднительно, можно лишь косвенно проследить их через вклад параметров $\varphi$ и $\omega$ в $g_1$ и $g_2$.

Оценивая ошибку для $g_1(n)$ и $g_2(n)$ на каждом временном шаге $n$, можно корректировать $u_1(n)$ и $u_2(n)$, корректируя $\varphi$ и $\omega$. Для корректировки можно использовать метод стохастического градиента (градиентного спуска) [23]. Тогда задача получения оптимальной траектории сведется к поиску оптимальных составляющих сигналов управления.

Предполагается, что данную динамическую систему на основе сеточной модели нестационарной теплопроводности можно будет «обучить» и использовать для этого специальный набор данных, полученный из реальных систем управления тепловым агрегатом. В рамках обучения предполагается придерживаться классического предположения о разграничении выборок на обучающую, тестовую и валидационную.

Рассмотрим метод градиентного спуска для параметра $\varphi$, полагая, что для параметра $\omega$ вывод будет аналогичен. Согласно методу градиентного спуска, изменение параметра будет иметь вид:

$$(30) \quad \begin{cases} \Delta\varphi_y^{(n)} = -\eta \cdot \dfrac{\partial E^{(n)}}{\partial \varphi_y^{(n)}} \\[3mm] \Delta\varphi_x^{(n)} = -\eta \cdot \dfrac{\partial E^{(n)}}{\partial \varphi_x^{(n)}} \end{cases}$$

Производная целевой функции по $\varphi$ для $Ox$ будет иметь вид (31), а для $Oy$ – (32).

$$(31) \quad \frac{\partial E_x^{(n)}}{\partial \varphi_x} = \frac{\partial g_1(T_{x_k}^{(n)}, u_1(n))}{\partial \varphi_x} \cdot (y - g_1(T_{x_k}^{(n)}, u_1(n)))$$

$$(32) \quad \frac{\partial E_y^{(n)}}{\partial \varphi_y^{(n)}} = \frac{\partial g_2(T_{y_q}^{(n)}, u_2(n))}{\partial \varphi_y^{(n)}} \cdot (y - g_2(T_{y_q}^{(n)}, u_2(n)))$$

Предположим, что известно эталонное значение только конечного состояния системы (температура заготовки после ее выпуска из печи), а промежуточные состояния неизвестны. Исходя из принципа оптимальности, условный минимум на каждом шаге может быть вычислен только если известно значение целевой функции, а целевая функция не определена в явном виде для состояний между $t_0$ и $t_N$.

Для решения поставленной проблемы рассмотрим обратный «прогон», т.е. движение по состояниям системы от последнего к первому. Рассмотрим систему $Ox$, предполагая, что для $Oy$ вывод аналогичен. Ошибка (31) для последнего состояния будет рассчитываться при $n = N$.

Для того, чтобы получить ошибку на шаге $t_{N-1}$, нужно определить производную целевой функции на шаге $t_{N-1}$, а она на нем определена неявно. Для того, чтобы рассчитать функцию в неявном виде, необходимо обратиться к уравнению (23) для изменения состояния системы. Отсюда следует, что вклад каждого предыдущего состояния системы в будущие является мультипликативным с некоторым коэффициентом $\delta_n$:

$$(33) \quad \xi_{x_k}^{(N)} = \prod_{n=0}^{N-1} \delta_n \xi_{x_k}^{(n)}$$

Коэффициент $\delta_n$ на каждом шаге $n$ может быть вычислен в виде:

$$(34) \quad \delta_n = \frac{h_x^2 \cdot \varphi_x^{(n)}}{\varphi_x^{(n)} \cdot h_x^2 + 2 \cdot (\frac{\varphi_x^{(n)}}{\omega_x^{(n)}}) \cdot \tau \cdot \varphi_x^{(n)} \cdot (1 - \alpha_{x_{k-1}})}$$

Тогда, согласно (33), ошибка тоже будет накапливаться мультипликативно:

$$(35) \quad E_x^{(N)} = \hat{y} - (\prod_{n=0}^{N-1} \delta_n \xi_{x_k}^{(n)}), k = \overline{1, H-1}$$

В результате, согласно принципа оптимальности, можно вывести условный минимум функции ошибки на $N$-1-м шаге:

$$(36) \quad E^{(N-1)} = \min_{u_1(N-1)} [\delta_{N-1} \cdot g_1(T_{x_k}^{(N-1)}, u_1(N-1)) \cdot E^{(N)}]$$

Заметим, что для нахождения производной данной функции ошибки на $N$-1-м шаге по $\varphi$ нужно знать ошибку на предыдущем шаге и производную функции состояния по управлению и по самому состоянию:

$$(37) \quad \frac{\partial E_x^{(N-1)}}{\partial \varphi_x} = \delta_{N-1} \cdot (\frac{\partial g_1(T_{x_k}^{(N-1)}, u_1(N-1))}{\partial T_x^{(N)}} \cdot \frac{\partial g_1(T_{x_k}^{(N)}, u_1(N))}{\partial \varphi_x}) \cdot$$
$$\cdot (y - g_1(T_{x_k}^{(N)}, u_1(N))$$

Выводы (33)-(37) справедливы и для системы $Oy$, за исключением того, что уравнение (34) будет иметь иной вид в соответствии с уравнением состояния (26):

$$(38) \quad \begin{cases} \delta_n^{(y)} = \dfrac{\omega_y^{(n)} \cdot h_y^2}{X_n} - \sqrt[4]{\dfrac{2 \cdot \tau \cdot \varepsilon_2 \cdot \sigma \cdot h_y}{X_n}} \\ X_n = 2 \cdot \tau \cdot \varphi_y^{(n)} \cdot (1 - \alpha_{y_q}^{(n-1)}) + 2 \cdot \tau \cdot \kappa_2 \cdot h_y + \omega_y^{(n)} \cdot h_y^2 \end{cases}$$

И, соответственно, уравнение ошибки для системы $Oy$ будет иметь вид:

$$(39) \quad \frac{\partial E_y^{(N-1)}}{\partial \varphi_y} = \delta_{N-1}^{(y)} \cdot (\frac{\partial g_2(T_{y_q}^{(N-1)}, u_2(N-1))}{\partial T_y^{(N)}} \cdot \frac{\partial g_2(T_y^{(N)}, u_2(N))}{\partial \varphi_y} \cdot$$
$$\cdot (y - g_2(T_{y_q}^{(N)}, u_2(N))))$$

Определим частные производные уравнений состояний $g_1$ и $g_2$ по самому состоянию в момент времени $N$:

$$(40) \begin{cases} \dfrac{\partial g_1}{\partial T_{x_k}^{(N)}} = \dfrac{h_x^2}{2 \cdot (\dfrac{\varphi_x^{(n)} \cdot \tau \cdot (1 - \alpha_{y_k}^{(n-1)})}{\omega_x^{(n)}}) + h_x^2} \\[4mm] \dfrac{\partial g_2}{\partial T_{y_k}^{(N)}} = \dfrac{h_y \cdot (h_y \cdot \omega_y^{(n)} - 8 \cdot (\varepsilon_2 \cdot \sigma \cdot \tau \cdot (T_{y_k}^{(n)})^3))}{h_y^2 \cdot \omega_y^{(n)} + \tau \cdot (2 \cdot (\varphi_y^{(n)} \cdot (1 - \alpha_{y_k}^{(n-1)})) + 2 \cdot (h_y \cdot \kappa_2))} \end{cases}$$

Для определения частной производной $g_1$ по $\varphi_x$ воспользуемся формулой состояния (25) и для упрощения расчета производной сложной функции сделаем следующие замены повторяющихся частей функции:

$$(41) \begin{cases} x1 = h_x^2 \\[2mm] x2 = \varphi_x^{(n)} \cdot \tau \cdot \dfrac{(1 - \alpha_{x_k}^{(n-1)})}{\omega_x^{(n)}} \\[2mm] x3 = 2 \cdot (h_x \cdot q_2) \\[2mm] x4 = \beta_{x_k}^{(n-1)} \cdot \varphi_x^{(n)} \\[2mm] x5 = \varphi_x^{(n)} \cdot (2 \cdot (\dfrac{\varphi_x^{(n)} \cdot \tau \cdot (1 - \alpha_{x_k}^{(n-1)})}{\omega_x^{(n)}}) + h_x^2) \\[2mm] x6 = h_x^2 \cdot T_{x_k}^{(n)} \end{cases}$$

Тогда частная производная будет иметь вид:

$$\frac{\partial g_1}{\partial \varphi_x^{(n)}} = \frac{(x6 + \dfrac{\tau \cdot (4 \cdot x4 - x3)}{\omega_x^{(n)}}) \cdot x6^2}{x5 - \varphi_x^{(n)} \cdot (4 \cdot (x2 + x1)) \cdot (x6 + \dfrac{\tau \cdot (2 \cdot x4 - x3)}{\omega_x^{(n)}})}$$

(42)

Для $g_2$ по $\varphi_y$ воспользуемся уравнением (26) и выполним замену:

$$(43) \begin{cases} x1 = 1 - \alpha_{y_k}^{(n-1)} \\[2mm] x2 = h_y \cdot \kappa_2 \\[2mm] x3 = h_y^2 \cdot \omega_y^{(n)} + \tau \cdot (2 \cdot (\varphi_y^{(n)} \cdot (1 - \alpha_{y_q}^{(n-1)})) + 2 \cdot h_y \cdot \kappa_2 \end{cases}$$

В результате получим уравнение частной производной вида:

$$(44) \quad \frac{\partial g_2}{\partial \varphi_y^{(n)}} = \tau \cdot (2 \cdot \beta_{y_q}^{(n-1)} - 2 \cdot (x1 \cdot (\frac{h_y \cdot (2 \cdot (x2 \cdot \sigma \cdot \tau \cdot (T_2^4 - (T_{y_q}^n)^4)) + h_y \cdot \omega_y^{(n)} \cdot T_{y_q}^n)}{x3} +$$

$$+ \frac{\tau \cdot (2 \cdot (\beta_{y_q}^{(n-1)} \cdot \varphi_y^{(n)}) + 2 \cdot (x2 \cdot T_{y_q}^n))}{x3})) \cdot \frac{1}{x3}$$

Аналогичным образом, найдем частные производные $g_1$ и $g_2$ по $\omega$. Для начала получим формулу расчета частной производной уравнения состояния $g_1$ по $\omega_x$, и в результате получим систему (45) с произведенными заменами для упрощения восприятия формулы и более удобного использования в численном эксперименте.

$$(45) \quad \begin{cases} \frac{\partial g_1}{\partial \omega_x^{(n)}} = \tau \cdot (\frac{(x5 - x3)}{x4} + 2 \cdot (\frac{(\varphi_x^{(n)})^3 \cdot x1 \cdot (x2 \cdot T_{x_k}^{(n)} + \tau \cdot \frac{(x3 - x5)}{\omega_x^{(n)}})}{(\varphi_x^{(n)} \cdot x4)^2})) \cdot \frac{1}{(\omega_x^{(n)})^2} \\ x1 = 1 - \alpha_{x_k}^{(n-1)} \\ x2 = h_x^2 \\ x3 = 2 \cdot (\beta_{x_k}^{(n-1)} \cdot \varphi_x^{(n)}) \\ x4 = 2 \cdot (\frac{\varphi_x^{(n)} \cdot \tau \cdot (1 - \alpha_{x_k}^{(n-1)})}{\omega_x^{(n)}}) + h_x^2 \\ x5 = 2 \cdot h_x \cdot q_2 \end{cases}$$

Аналогичным образом получим частную производную $g_2$ по $\omega_y$ в виде системы с заменами:

$$(46) \quad \begin{cases} \frac{\partial g_2}{\partial \omega_y^{(n)}} = x2 \cdot (T_{y_q}^{(n)} - \frac{(h_y \cdot (2 \cdot (x2 \cdot \sigma \cdot \tau \cdot (T_2^4 - (T_{y_q}^{(n)})^4)) + h_y \cdot \omega_y^{(n)} \cdot T_{y_q}^{(n)})}{x3}) + \\ + \frac{\tau \cdot (2 \cdot (\beta_{y_q}^{(n-1)} \cdot \varphi_y^{(n)}) + 2 \cdot (x1 \cdot T_2)))}{x3} \cdot \frac{1}{x3} \\ \\ x1 = h_y \cdot \kappa_2 \\ x2 = h_y^2 \\ x3 = h_y^2 \cdot \omega_y^{(n)} + \tau \cdot (2 \cdot (\varphi_y^{(n)} \cdot (1 - \alpha_{y_q}^{(n-1)})) + 2 \cdot h_y \cdot \kappa_2 \end{cases}$$

Таким образом, используя полученные выше формулы для частных производных, можно для каждого $n \neq N$ вычислять ошибку вида (37) для каждой отдельной динамической системы ($Ox$ и $Oy$) и проводить коррекцию соответствующих коэффициентов $\varphi^{(n)}$ и $\omega^{(n)}$, полагая движение в сторону оптимальной траектории согласно функциональному уравнению (36). Суть данного процесса может быть проиллюстрирована следующей схемой (рис. 3).

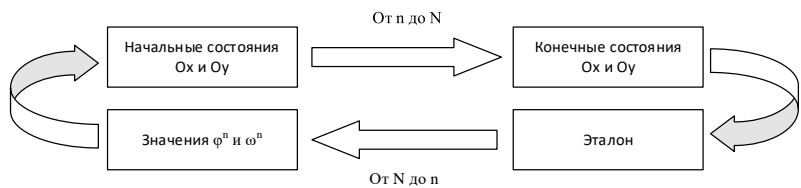

*Рис. 3. Принцип «обучения» сеточной модели*

## 5. Численные эксперименты

Для проведение численного эксперимента было решено реализовать собственную среду машинного обучения на языке C#. От использования готовых сред моделирования было решено отказаться, так как необходимо было реализовать свой подход, а не воспользоваться существующим.

В результате разработки на базе IDE Visual Studio 2022 была разработана программа для ЭВМ. В ядре представленной программы находится сеточная модель нестационарной теплопроводности в виде (22) с соответствующими заменами теплофизических коэффициентов вида (19) в уравнениях (10), (13), (14), (16) и (17).

В качестве исходных данных для «обучения» сеточной модели использовались данные, полученные из действующих систем слежения за металлом в методических печах нагрева металла перед прокатом одного из металлургических комбинатов Белгородской области. Поскольку рассматривалась шестизонная нагревательная печь, то история нагрева была собрана по зонам. Размер выборки для группы углеродистых сталей, обладающих сходными физико-химическими характеристиками, составил 6098 записей о заготовках, разбитых по следующим признакам:
1. Время нагрева в первой и второй зонах печи;
2. Время нагрева в третьей и четвертой зонах печи;

3. Время нагрева в пятой и шестой зонах печи;
4. Температуры в первой, третьей и пятой зонах печи;
5. Температуры во второй, четвертой и шестой зонах печи.
6. Давление в первой и второй зонах печи;
7. Давление в третьей и четвертой зонах печи;
8. Давление в пятой и шестой зонах печи;
9. Время остывания заготовки при движении до точки, где установлен пирометр, измеряющий результирующую температуру;
10. Сама результирующая температура, снятая пирометром, используемая в качестве эталона при обучении.

Исходная выборка была разбита на обучающее множество (80 % от всех записей) и тестовое множество (20 % от всех записей). Обучение модели происходила порциями по 550 эпох обучения.

Для того, чтобы избежать переобучения, было решено добавить в модель L1-регуляризацию (LASSO-регуляризацию), имеющую следующий вид:

$$(47)\quad L = \frac{1}{2}\sum_{i=1}^{M}(y_i - g_1(T_{x_k}^{(n)}, u_1(n)_i) + \lambda_1 \left\| u_1(n) \right\|_1 ,$$

где, $L$ – это целевая функция с учетом регуляризации. Основная идея L1-регуляризации заключается во ведении штрафа на величину настраиваемых коэффициентов [20]. Значение параметра $\lambda_1$ является гиперпараметром и подбирается экспериментально. В результате тестовых прогонов было решено использовать $\lambda_1 = 0{,}05$.

В качестве метрики оценки качества модели было решено использовать абсолютное среднее:

$$(48)\quad MAE = \frac{\sum_{i=1}^{m}(y_i - g_1(T_{x_k}^{(n)}, u_1(n)_i)}{m} ,$$

где m – это размер тестового множества ($m \in M$). В рамках всего численного эксперимента использовалась постоянная сетка для прямого решения модели нестационарной теплопроводности размерностью *100х100* ($H = 100$, $W = 100$) по пространству и *50х1* ($N$=50) по времени. Также было принято решение оставить шаг обучения постоянным значением ($\eta = 0{,}001$). После каждой эпохи обучения модель получала на вход данные из тестового множества, на основе которых оценивалась метрика (48). Критерием остановки служило замедление градиента ошибки ($\nabla e < 10^{-6}$).

На рисунке 4 представлены 550 эпох обучения модели, где по оси $Oy$ отложены значение метрики (48) для тестового множества. Как можно заметить, инициализация параметров генератором псевдослучайных положительных чисел привела к базовой ошибке 45,5 °С в среднем по всей выборке.

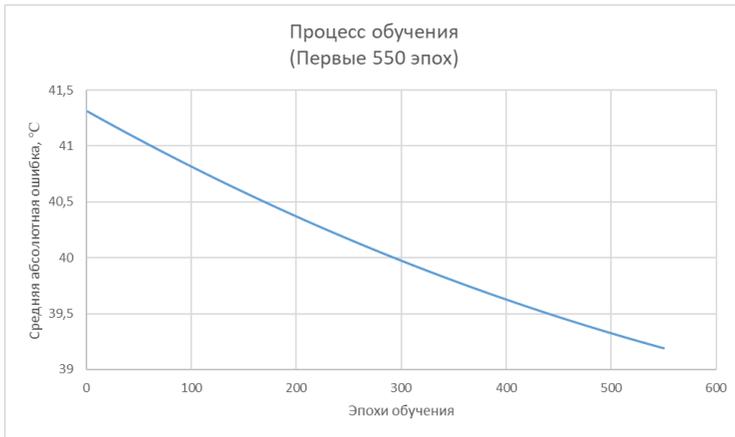

*Рис. 4– Средняя ошибка на первых 550 эпохах обучения*

За 550 эпох обучения ошибку удалось снизить в среднем на 2,11 °С. Предполагается, что подобный результат является следствием постоянного малого шага обучения. В результате, был проведен полный цикл обучения до срабатывания условия останова (рис. 5). Длительность обучения составила 8450 эпох, в среднем по 2,4 секунды на одну эпоху, что равняется приблизительно 5,7 часам обучения.

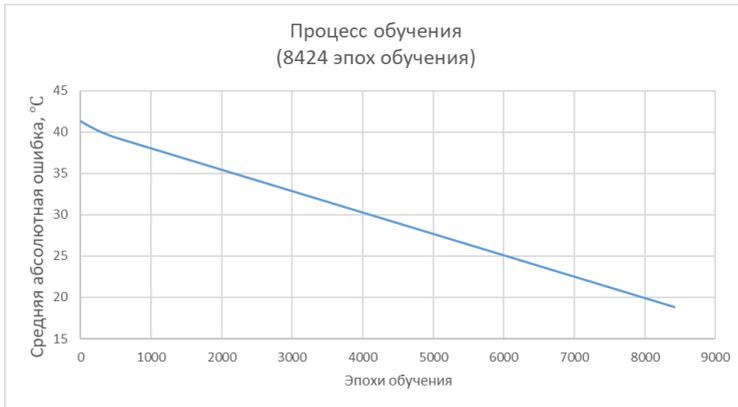

*Рис. 5 – Полный цикл обучения до срабатывания условия остановки*

За полный цикл обучения удалось снизить ошибку на 22,48 ℃ в среднем по всей тестовой выборке. В результате эксперимента было установлено, что имеет место ограничение модели вида:

(49) $N > H \cdot W$.

Несоблюдение условия (49) приводит к нарушению условия $\omega \neq 0$ при достаточно больших $\varphi$. Также, важно отметить, что в процессе численного эксперимента первичная инициализация параметров $\varphi$ и $\omega$ проводилось случайным образом на положительной числовой оси.

В процессе проведения численного эксперимента было установлено, что модель может быть усовершенствована в будущем, в частности, путем добавления адаптивного шага обучения. Тем не менее, имеющаяся на данный момент модель позволила получить ошибку на тестовом множестве в среднем 18,82 ℃. Полученная модель незначительно хуже модели, адаптированной более классическом способом (регрессионными уравнениями по той же группе схожих по физико-химических конфигурациям нагреваемого вещества сталей), имеющую точность в 18,1 ℃ [11].

Модель, обученная предложенным методом, не получала никаких сведений о физико-химической структуре нагреваемого вещества явно, но смогла достичь ошибки не хуже адаптированной модели, восстановив температурные зависимости плотности, теплоемкости и теплопроводности неявным образом. Таким образом, если с реальной печи будут получены технологические данные по группе марок сталей, по которым отсутствуют табличные данные, то предлагаемый подход позволит получить адекватную модель нагрева таких заготовок. При этом классический метод адаптации окажется неприменим в данной ситуации ввиду отсутствия уже упомянутых табличных данных.

Предполагается, что доработка модели позволит улучшить имеющийся результат, увеличить скорость обучения и повысить устойчивость самого процесса градиентного спуска.

## *6. Заключение*

Полученные в работе формулы позволяют «обучать» сеточную модель, используя в качестве входных данных информацию с конкретных производств и технологических объектов, что сделает итоговую модель более конкретной по отношению к моделируемому контексту. Другими словами, сеточная модель получит преимущество моделей машинного обучения в естественной адаптации к нагреваемому объекту.

Наличие естественной адаптации сделает модель более практически применимой, так как она сможет обобщать скрытые зависимости неизвестного вида, находящиеся в технологических данных. Данный тезис проиллюстрирован в численном эксперименте, по результатам которого модель смогла достичь среднего уровня ошибки, сопоставимого с уровнем классически адаптированной модели, не получая никаких сведений о структуре нагреваемого вещества в явном виде.

Таким образом, процесс поиска температурных зависимостей становится частью жизненного цикла модели. Данный факт позволяет исключить из этапа построения модели те процедуры, которые связанные с адаптацией модели к нагреваемому веществу. Это позволит применять её в ситуациях, когда адаптировать теплофизические зависимости явным образом (например, регрессиями) не представляется возможным. Например, при наличии множества различных физико-химических конфигураций, или для таких конфигурации, у которых не имеется в достаточном количестве дискретных экспериментальных замеров, для которых можно было бы применить инструментарий регрессионных моделей.

## 7. Литература

# IMPLICIT ADAPTATION OF MESH MODEL OF TRANSIENT HEAT CONDUCTION PROBLEM

**Zhukov Petr**, STI NUST "MISIS", Stary Oskol, postgraduate student (Zhukov.petr86@yandex.ru).

**Glushchenko Anton**, ICS RAS, Moscow, Doctor of Sciences, docent (aiglush@ipu.ru).

**Fomin Andrey,** STI NUST "MISIS", Stary Oskol, Candidate of Technical Sciences, docent

Abstract: Considering high-temperature heating, the equations of transient heat conduction model require an adaptation, i.e. the dependence of thermophysical parameters of the model on the temperature is to be identified for each specific material to be heated. This problem is most often solved by approximation of the tabular data on the measurements of the required parameters, which can be found in the literature, by means of regression equations. But, for example, considering the steel heating process, this approach is difficult to be implemented due to the lack of tabular discrete measurements for many grades of steel, such as alloyed ones. In this paper, the new approach is proposed, which is based on a solution of a related variational problem. Its main idea is to substitute the adaptation process in the classical sense (i.e., to find the dependencies of thermophysical parameters on temperature) with "supervised learning" of a mesh model on the basis of the technological data received from the plant. The equations to adjust the parameters of the transient heat conduction model, which are related to the thermophysical coefficients, have been derived. A numerical experiment is conducted for steel of a particular group of grades, for which enough both technological as well as tabular data are available. As a result, the "trained" mesh model, which has not received explicitly any information about the physical and chemical properties of the heated substance, demonstrated an average error of 18.82 0C, which is quite close to the average error of the model adapted classically on the basis of the tabular data (18.1 0C).

Keywords: mesh model, stochastic model of transient heat conduction; model adaptation; gradient descent.

УДК 536.3 + 519.6
ББК 22.193